\documentstyle[aps,preprint,tighten,psfig]{revtex}

\newcommand{\lsim}{\mathrel{\mathop{\kern 0pt \rlap
  {\raise.2ex\hbox{$<$}}}
  \lower.9ex\hbox{\kern-.190em $\sim$}}}
\newcommand{\gsim}{\mathrel{\mathop{\kern 0pt \rlap
  {\raise.2ex\hbox{$>$}}}
  \lower.9ex\hbox{\kern-.190em $\sim$}}}
\newcommand{\beq}    {\begin{equation}}
\newcommand{\eeq}    {\end{equation}}
\newcommand{\beqarr} {\begin{eqnarray}}
\newcommand{\eeqarr} {\end{eqnarray}}
\newcommand{\barr}   {\begin{array}}
\newcommand{\earr}   {\end{array}}

\begin{document}

\preprint{
\begin{tabular}{r}
DFTT 18/99 \\
ROM2F/99/07 \\
LAPTH--727/99 \\
FTUV/99--22 \\
IFIC/99--23
\end{tabular}
}

\title{Extending the DAMA annual--modulation region by inclusion of 
         the uncertainties in astrophysical velocities}

\author{\bf 
P. Belli$^{\mbox{a}}$
\footnote{E--mail: belli@roma2.infn.it, bernabei@roma2.infn.it,
bottino@to.infn.it, donato@lapp.in2p3.fr, prosperi@roma1.infn.it,
fornengo@flamenco.ific.uv.es, scopel@posta.unizar.es},
R. Bernabei$^{\mbox{a}}$,
A. Bottino$^{\mbox{b}}$,
F. Donato$^{\mbox{c}}$, \\
N. Fornengo$^{\mbox{d}}$,
D. Prosperi$^{\mbox{e}}$,
S. Scopel$^{\mbox{f}}$\footnote{INFN Post--doctoral Fellow}
\vspace{6mm}
}

\address{
\begin{tabular}{c}
$^{\mbox{a}}$
Dipartimento di Fisica,  Universit\`a di Roma ``Tor Vergata'' \\
and INFN, Sez. di Roma2, I--00133 Roma, Italy\\
$^{\mbox{b}}$
Dipartimento di Fisica Teorica, Universit\`a di Torino \\
and INFN, Sez. di Torino, Via P. Giuria 1, I--10125 Torino, Italy\\
$^{\mbox{c}}$
Laboratoire de Physique  Th\'eorique LAPTH, B.P. 110, F--74941\\
Annecy--le--Vieux Cedex, France \\
and INFN, Sez. di Torino, Via P. Giuria 1, I--10125 Torino, Italy\\
$^{\mbox{d}}$
Instituto de F\'{\i}sica Corpuscular -- C.S.I.C. --
Departamento de F\'{\i}sica Te\`orica, \\
Universitat de Val\`encia, E-46100 Burjassot, Val\`encia, Spain \\
$^{\mbox{e}}$
Dipartimento di Fisica,  Universit\`a di Roma ``La Sapienza'' \\
and INFN, Sez. di Roma, I--00185,  Roma, Italy\\
$^{\mbox{f}}$ 
Instituto de F\'\i sica Nuclear y Altas Energ\'\i as, 
Facultad de Ciencias, \\
Universidad de Zaragoza, Plaza de San Francisco s/n, E--50009 Zaragoza, Spain
\end{tabular}
}

\maketitle

\begin{abstract}

The original annual--modulation region, singled out by the 
DAMA/NaI experiment for direct detection of WIMPs, 
is extended by taking into account the uncertainties in the galactic 
astrophysical velocities. Also the effect due to a possible 
bulk rotation for the dark matter halo is considered. 
   We find that the range for the WIMP 
mass becomes 30 GeV $\lsim m_{\chi} \lsim$ 130 GeV at 1--$\sigma$ C.L. 
with a further
extension in the upper bound, when a possible bulk rotation of the dark matter 
halo is taken into account. 
We show that the DAMA results, when interpreted in the framework of 
the Minimal Supersymmetric extension of the Standard Model, 
are consistent with 
a relic neutralino as a dominant component of cold dark matter (on the 
average in our universe and in our galactic halo). 
It is also discussed the discovery potential for the relevant 
supersymmetric configurations at accelerators of present generation. 
 
\end{abstract}  


\section{Introduction}

   The DAMA/NaI Collaboration has recently reported the indication of an 
annual--modulation effect in a direct search experiment for WIMPs 
\cite{dama1,dama2,dama3}. In Ref. \cite{dama3} it has been shown that 
a statistical (maximum--likelihood) analysis of the data concerning to a 
total exposure of  19,511 kg $\times$ day provides, 
at a 2--$\sigma$ C.L., a well delimited region in the plane 
$\rho_{\chi} \sigma^{(\rm nucleon)}_{\rm scalar}$ -- $m_\chi$, 
where $m_\chi$ is the WIMP mass, $\rho_{\chi}$ is its local (solar
neighbourhood) density and 
$\sigma^{(\rm nucleon)}_{\rm scalar}$ is the WIMP--nucleon  scalar elastic
cross section. Obviously, the location 
of the annual--modulation region depends on the functional form adopted 
for the speed distribution of 
the dark matter particles, and  on the values assigned to the galactic 
astrophysical velocities. In Refs. \cite{dama1,dama2,dama3} a
standard Maxwellian distribution was taken for the WIMP speed distribution 
with a root mean square velocity 
$v_{\rm rms}$ = 270  Km s$^{-1}$ and a WIMP escape velocity in the halo 
$v_{\rm esc} = 650$ Km s$^{-1}$. 

   The DAMA/NaI results on annual--modulation were analyzed in Refs. 
\cite{noi1,noi2,noi3,noi4,noi5} in terms of relic neutralinos. It was 
proved that the DAMA/NaI data are compatible with a neutralino as a major 
component of dark matter in the universe. In 
Refs. \cite{noi1,noi2,noi3,noi4,noi5} 
it was also   
shown that a significant number of the relevant supersymmetric 
configurations are compatible with supergravity schemes 
\cite{an}, are explorable at accelerators and/or by indirect searches 
for relic particles 
(up-going muons at neutrino telescopes and antiprotons in space). 

   In the present paper we discuss: i) the extension  of 
the DAMA/NaI annual--modulation region in the plane 
$\rho_{\chi} \sigma^{(\rm nucleon)}_{\rm scalar}$ -- $m_\chi$, 
when the uncertainties of the galactic astrophysical velocities are taken into 
account,  
and ii) the ensuing consequences of this extension in terms of 
properties for relic neutralinos and for the related supersymmetric 
configurations \cite{prel}. 

\section{Extended annual--modulation region}   

The differential rate for WIMP  direct detection is given by

\begin{equation}
\frac {dR}{dE_R}=N_{T}\frac{\rho_{\chi}}{m_{\chi}}
                    \int_{v_{\rm min}}^{v_{\rm max}} \,d \vec{v}\,f(\vec v)\,v
                    \frac{d\sigma}{dE_{R}}(v,E_{R}) \, , 
\label{eq:diffrate0}
\end{equation}
where $N_T$ is the number of the target nuclei per unit of mass,
$\vec v$ and $f(\vec v)$ denote the WIMP
velocity and velocity distribution function  in the Earth frame
($v = |\vec v|$) and $d\sigma/dE_R$ is the WIMP--nucleus differential 
cross section.
The nuclear recoil energy is given by
$E_R={{m_{\rm red}^2}}v^2(1-\cos \theta^*)/{m_N}$,
where $\theta^*$ is the scattering
angle in the WIMP--nucleus center--of--mass frame,
$m_N$ is the nuclear mass and $m_{\rm red}$ is the WIMP--nucleus
reduced mass. The velocity $v_{\rm min}$ is defined as 
$v_{\rm min} = (m_N E_R/2m_{\rm red}^2)^{\frac{1}{2}}$ and 
$v_{\rm max}$ is the maximal 
(escape) WIMP velocity in the Earth reference frame.  
Eq.(\ref{eq:diffrate0}) refers to a mono-atomic
detector; the generalization to our case of NaI is straightforward.

In principle, the differential WIMP--nucleus cross section  is composed of a 
coherent part and a 
spin--dependent one. Making the safe assumption that the former one is 
dominant over the latter, we may write 

\begin{equation}
\frac{d \sigma}{d E_R}=\frac{\sigma_0}{E_R^{\rm max}} F^2(q) \, , 
\label{eq:dsigma_dq}
\end{equation}
where $\sigma_0$ is the point-like scalar WIMP--nucleus cross section, 
$E_R^{\rm max}$
is the maximum value of $E_R$ and  $F(q)$ denotes the nuclear form factor, 
expressed as a function
of the momentum transfer $q^2 \equiv\mid{\vec {q}}\mid^2=2m_NE_R$. 
$\sigma_0$ may be conveniently rewritten in terms of 
the scalar WIMP--nucleon cross section 
$\sigma^{(\rm nucleon)}_{\rm scalar}$

\beq
\sigma_0 = \left(\frac{1 + m_{\chi}/m_p}{1 + m_{\chi}/m_N} \right)^2 
\; A^2 \; \sigma^{(\rm nucleon)}_{\rm scalar}, 
\eeq 

\noindent
where $m_p$ is the proton mass and $A$ is the mass number of the nucleus. 

For the nuclear form factor in Eq.(\ref{eq:dsigma_dq}) 
we use the Helm parameterization of
the scalar form factor \cite{helm,engel}:
\beq
F(q)=3 \frac {j_1(qr_0)}  {qr_0} \exp \left(-\frac{1}{2} s^2 q^2 \right)
\label{eq:ff}
\eeq
\noindent
where $s \simeq 1$ fm is the
thickness parameter for
the nucleus surface, $r_0 = (r^2-5s^2)^{1/2}$, $r=1.2~A^{1/3}$ fm 
and
$j_1(qr_0)$ is the spherical Bessel function of index 1.

Once a functional form for $f(v)$ is chosen and specific values 
are assigned to the relevant astrophysical velocities, 
the quantity $\rho_{\chi} \sigma^{(\rm nucleon)}_{\rm scalar}$ 
may be extracted from  measurements of the differential rate 
$dR/dE_R$ as a function of $m_{\chi}$.  

Here, for $f(v)$ we take the standard Maxwell--Boltzmann distribution 
(as in an isothermal spherical model for the halo) with a finite escape 
velocity, i.e. in the galactic rest frame we write

\begin{equation}
f_{\rm gal}(v^{\rm gal}) = N  \left(\frac{3}{2 \pi v_{\rm rms}^{2}}\right)^{3/2}
\exp \left(- \frac{3 (v^{\rm gal})^2}{2 v_{\rm rms}^2}\right)  \, , 
\label{eq:MB}
\end{equation}
where the normalization factor is
\begin{equation}
N = \left[ \mbox{erf} (z)-\frac{2}{\sqrt{\pi}}
z \exp(-z^{2}) \right]^{-1}  \, , 
\label{eq:MBnorm}
\end{equation}
with $z^2 = 3v_{\rm esc}^2/(2v_{\rm rms}^2)$. 
In the isothermal spherical model, $v_{\rm rms}$ 
is related to the asymptotic value
$v_\infty$ of the rotational velocities by the relation
$v_{\rm rms} =\sqrt{\frac{3}{2}}v_\infty $.
The measured rotational velocity 
remains almost flat (to roughly 15\%) between 4 Kpc and 18 Kpc; here, 
we identify $v_\infty$ with the rotational velocity 
of the Local System at the position of the
Solar System  $v(r_{\odot}) \equiv v_0$, whose physical range will be shortly 
discussed.

To make use of Eq.(\ref{eq:diffrate0}), 
Eq.(\ref{eq:MB}) has to be transformed to the rest frame of the Earth,
which moves through the Galaxy with
a velocity  
\beq
v_\oplus = v_\odot + v_{\rm orb} \; \cos \gamma \; \cos[\omega (t - t_0)], 
\eeq
in the azimuthal direction. $v_\odot$ is given  by 

\beq
v_\odot = v_0 + 12 \; \mbox{km s}^{-1}. 
\eeq
In these expressions the speed of 12 km s$^{-1}$ stands for the motion 
of the Solar System with respect to the Local System, 
$v_{\rm orb} = 30$ km s$^{-1}$ denotes 
the Earth orbital speed around the Sun, the angle 
$\gamma \simeq 60^0$ is the inclination of the Earth orbital plane 
with respect to the galactic plane and $\omega = 2 \pi / 365 \mbox{days}$, 
$t_0 =$ June  2$^{\rm nd}$ \cite{dfs,ffg}.  

In Refs. \cite{dama1,dama2,dama3} the DAMA/NaI annual--modulation 
region was extracted from the experimental data, by using a maximum likelihood 
method and by taking  
for the velocities $v_0$ and $v_{\rm esc}$ the following values:  
$v_0 = 220$ km s$^{-1}$ and $v_{\rm esc} = 650$ km s$^{-1}$ (the relevant 
region is one of the domains displayed in Fig. 1). The ensuing 
implications for relic neutralinos, derived in 
 \cite{noi1,noi2,noi3,noi4,noi5},  refer to the same set of 
astrophysical velocities.  

In the present paper we extend the analysis 
of the annual--modulation region, by considering the  physical
ranges associated to $v_0$ and $v_{\rm esc}$ \cite{lt,k,c}:

\beq
v_0 = (220 \pm 50) \; \mbox{km s}^{-1} \;\; 
(90\% \; \mbox{C.L.})\, ,
\label{eq:not}
\eeq

\beq
v_{\rm esc} = (450 \div 650) \;  \mbox{km s}^{-1} \;\; 
(90\% \; \mbox{C.L.})\, .
\label{eq:esc}
\eeq

The statistical method for the extraction of the annual--modulation region is
the same as the one employed in Refs. \cite{dama1,dama2,dama3}, 
with a lower bound $m_{\chi} \geq$ 25 GeV. 

On the basis of the analytical properties of the time--modulated part 
of the detection rate \cite{ffg}, one expects that a variation in $v_0$ 
induces a sizeable modification in the range of $m_{\chi}$, without affecting 
$\rho_{\chi} \sigma^{(\rm nucleon)}_{\rm scalar}$ significantly. 
Indeed, an increase (decrease) in $v_0$ is expected to extend the 
original DAMA/NaI region toward lower (larger) WIMP masses for kinematical
reasons; $\rho_{\chi} \sigma^{(\rm nucleon)}_{\rm scalar}$ cannot appreciably change, 
since it acts as a normalization factor which is determined by the size of the
average detection rate. 
Also, no sizeable variation  of both $m_{\chi}$ and 
$\rho_{\chi} \sigma^{(\rm nucleon)}_{\rm scalar}$ is expected from  a variation in
$v_{\rm esc}$, since the escape velocity provides a cut--off in the integral of 
Eq.(\ref{eq:diffrate0})  applying only on the flat tail of the velocity
distribution $f(v)$. 

We report in Fig. 1 the (2--$\sigma$ C.L.) annual--modulation 
regions extracted from the 
DAMA/NaI data of Ref. \cite{dama3} (total exposure of 19,511 kg $\times$ day) 
for sets of values for $v_0$ and $v_{\rm esc}$,  which bracket the ranges given 
in Eqs.(\ref{eq:not} -- \ref{eq:esc}).  
The 1--$\sigma$ ranges for the relevant quantities 
are given in Table 1. Notice that in Fig. 1 the density $\rho_{\chi}$ is given
in units of 0.3 GeV cm$^{-3}$, i.e.
$\rho_{\chi}^{0.3} \equiv \rho_{\chi}$/(0.3 GeV cm$^{-3}$). 

{}From the features displayed in Fig. 1 we notice that,
as anticipated above: 
 i) the location of the annual--modulation region is rather sensitive 
to the velocity $v_0$, with a sizeable extension in the range of 
$m_{\chi}$, but with small variations  in 
$\rho_{\chi} \sigma^{(\rm nucleon)}_{\rm scalar}$; 
ii) the annual--modulation region is  essentially independent of 
$v_{\rm esc}$. {}From the results of Table I we conclude that the uncertainties 
in 
$v_0$ extend the range of $m_{\chi}$, as singled out by the DAMA/NaI 
annual--modulation data, to (at 1--$\sigma$  C.L.)

\beq
30 \; \mbox{GeV} \lsim m_{\chi} \lsim 130 \; \mbox{GeV} \, .
\label{eq:rangem}
\eeq

Taking into account the whole ranges of $v_0$ and $v_{\rm esc}$ 
(see Eqs. (\ref{eq:not}-\ref{eq:esc})), we find the annual--modulation 
region which is 
depicted in Fig. 2. This region, which envelops the domains displayed in 
Fig. 1, will be hereafter referred to as region $R_m$. 

It is worth noticing that this annual--modulation region might be 
further extended, if a possible bulk rotation of the dark matter 
halo is introduced.  Unfortunately, halo models which take realistically 
into account this phenomenon are not yet available; 
however, effects of rotation of the dark matter halo on direct 
detection rates and ensuing upper bounds on cross sections have 
been addressed in some extreme models \cite{kk,rot}.
To obtain an estimate of the 
effects of a possible rotation of the
isothermal sphere one can consider a class of models \cite{lynden-bell2}
which describe the fastest rotating steady state by means of the following 
recipe:

\begin{eqnarray}
f_{+} (v^{\rm gal}) = \left \{
\begin{array}{ll}   f(v^{\rm gal}) \hspace{15mm} & v_\phi^{\rm gal} > 0 \\
                    0              & v_\phi^{\rm gal} < 0
\end{array}
\right.
\label{eq:f_piu}
\end{eqnarray}

\begin{eqnarray}
f_{-} (v^{\rm gal}) = \left \{
\begin{array}{ll}   0     \hspace{24mm}  & v_\phi^{\rm gal} > 0\\
                    f(v^{\rm gal})       & v_\phi^{\rm gal} < 0
\end{array}
\right.
\label{eq:f_meno}
\end{eqnarray}
where $v_\phi^{\rm gal}$ is the azimuthal component of $\vec v$.

Then one can  combine the above functions in order to deal with a more general 
family of distribution functions  $f_{\rm rot} (v^{\rm gal})$, 
defined as \cite{kk}:

\begin{equation}
f_{\rm rot}(v^{\rm gal}) = a f_{+} (v^{\rm gal}) + (1-a) f_{-} (v^{\rm gal}) \, , 
\label{eq:f_rot}
\end{equation}
where  $a$ is related to the dimensionless galactic angular momentum parameter 
$\lambda$ by the relation $\lambda = 0.36 |a-0.5|$.
In order to be consistent with 
the available extensive numerical work on galaxy formation, 
$\lambda$ should not exceed the value 0.05 \cite{lambda_gal}. 

Adopting the expression in Eq.(\ref{eq:f_rot}) for the velocity 
distribution function, one  finds that the
annual--modulation region may be extended to the domain depicted in Fig. 3, 
and 
the relevant range of $m_{\chi}$ would become (at 1--$\sigma$ C.L.)

\beq
30 \;  \mbox{GeV}  \lsim m_{\chi} \lsim 180 \; \mbox{GeV} \, .
\label{eq:rangem1}
\eeq

Although the possible occurrence of a bulk rotation of the dark matter halo 
is a quite interesting possibility deserving  further investigation, all 
subsequent analyses of the present paper will refer to the region $R_m$, 
given in Fig. 2.

\section{Supersymmetric Model}

The WIMP candidate considered in this paper is the neutralino, defined 
as the lowest--mass linear superposition of photino ($\tilde \gamma$),
zino ($\tilde Z$) and the two higgsino states
($\tilde H_1^{\circ}$, $\tilde H_2^{\circ}$) \cite{susy}
\begin{equation}
\chi \equiv a_1 \tilde \gamma + a_2 \tilde Z + a_3 \tilde H_1^{\circ}  
+ a_4 \tilde H_2^{\circ} \, . 
\label{eq:neu}
\end{equation}

To classify the nature of the neutralino it is useful to define a parameter 
$P \equiv a_1^2 + a_2^2$; hereafter the neutralino is called a {\it gaugino}, 
when $P > 0.9$, is called 
{\it mixed} when $0.1 \leq P \leq 0.9$ and a {\it higgsino} when $P < 0.1$. 

The theoretical framework adopted here is the Minimal Supersymmetric extension 
of the Standard Model (MSSM)\cite{susy}, which conveniently  describes the 
supersymmetric phenomenology at the electroweak scale, without being 
constrained by too strong theoretical assumptions. Specific details of the 
scheme employed in this paper are given in Ref. \cite{noi3}; here we only recall
a few essentials. 

The large number of free parameters inherent in the model 
is reduced to six independent ones, by
imposing a few assumptions at the electroweak scale: 
i) all trilinear parameters are set to zero except those of the third family, 
which are unified to a common value $A$;
ii) all squarks and sleptons soft--mass parameters are taken as 
degenerate: $m_{\tilde l_i} = m_{\tilde q_i} \equiv m_0$, 
iii) the gaugino masses are assumed to unify at $M_{GUT}$, and this implies that
the $U(1)$ and $SU(2)$ gaugino masses are related at the electroweak scale by 
$M_1= (5/3) \tan^2 \theta_W M_2$. 

 The six independent parameters are taken to be: 
$M_2, \mu, \tan\beta, m_A, m_0, A$, where ${\mu}$ is the Higgs--mixing 
parameter, $\tan \beta$ is the ratio of the two Higgs vacuum expectation 
values, and $m_A$ is the mass of the neutral pseudoscalar Higgs boson. To get
the scatter plots shown in this paper, the supersymmetric space has been 
scanned, by varying the parameters in
the following ranges: $10\;\mbox{GeV} \leq M_2 \leq  1000\;\mbox{GeV},\;
10\;\mbox{GeV} \leq |\mu| \leq  1000\;\mbox{\rm GeV},\;
80\;\mbox{GeV} \leq m_A \leq  1\;\mbox{TeV},\; 
100\;\mbox{GeV} \leq m_0 \leq  1\;\mbox{TeV},\;
-3 \leq A \leq +3,\;
1 \leq \tan \beta \leq 50$. 

Our supersymmetric parameter space is further constrained by
all the experimental limits obtained from accelerators on
supersymmetric and Higgs searches. The latest data from 
LEP2 on Higgs, neutralino, chargino and sfermion masses are 
used \cite{lep189}. Also the constraints 
due to the $b \rightarrow s + \gamma$ process \cite{glenn,barate} 
have been taken into account (see Ref. \cite{noi3} for the theoretical 
details).  

The supersymmetric parameter space has also been constrained by 
the requirement that the neutralino is the Lightest 
Supersymmetric Particle (LSP), i.e.  regions where the 
gluino or squarks or 
sleptons are lighter than the neutralino have been excluded. 

One further constraint is due to the requirement that the neutralino relic 
abundance does not exceed the cosmological bound, derivable from
measurements of the age of the Universe \cite{age} and 
of the Hubble constant \cite{hubble}. 
We have adopted here a conservative upper bound, 
$\Omega_{\chi}h^2 \leq 0.7$ 
($h$ is the usual Hubble parameter, defined in terms of the present--day 
value $H_0$ of the Hubble constant as 
$h \equiv H_0/(100~$ km$~$s$^{-1}~$Mpc$^{-1})$). 

The neutralino relic abundance is calculated here as in Ref.\cite{ouromega}. 
Inclusion of coannihilation effects \cite{coannih}
in the calculation of $\Omega_{\chi} h^2$ are not necessary here, 
since the instances under which these effects might be sizeable are 
marginal for the supersymmetric configurations concerning the DAMA/NaI data.

Whenever we have to evaluate the neutralino local density, we employ the
rescaling procedure. 
This rescaling consists in assuming that the neutralino local density 
$\rho_{\chi}$ may be taken as  $\rho_{\chi} = \xi \rho_l$ 
($\rho_l$ is the total local density of non-baryonic dark matter), 
with $\xi = \mbox{min}\;(1, \; \Omega_{\chi}  h^2 / (\Omega  h^2)_{\rm min})$, 
i.e., $\rho_{\chi}$ may be set equal to $\rho_l$ only when 
$\Omega_\chi h^2$ is larger than a minimal value
$(\Omega h^2)_{\rm min}$, compatible with observational data and with 
large--scale 
structure calculations; otherwise, 
when $\Omega_\chi h^2$ turns out  to be less than $(\Omega h^2)_{\rm min}$, 
and then the neutralino may only provide a fractional contribution
${\Omega_\chi h^2 / (\Omega h^2)_{\rm min}}$
to $\Omega h^2$, $\rho_{\chi}$ is reduced by the same fraction 
$\xi = {\Omega_\chi h^2 / (\Omega h^2)_{\rm min}}$ as compared to $\rho_l$. 
The value to be assigned to $(\Omega h^2)_{\rm min}$ is
somewhat arbitrary, in the range 
$0.01 \lsim (\Omega h^2)_{\rm min} \lsim 0.2$. In the present paper, whenever 
we have to apply rescaling, we use the value $(\Omega  h^2)_{\rm min} = 0.01$, 
which is conservatively derived from the
estimate $\Omega_{\rm galactic} \sim 0.03$. 

In Fig. 2 we display our results for 
$\rho_{\chi}^{0.3} \; \sigma^{(\rm nucleon)}_{\rm scalar}$ in the form of the 
scatter plot which is derived by scanning the supersymmetric 
parameter space over the grid
defined above. 
For the evaluation of $\sigma^{(\rm nucleon)}_{\rm scalar}$ we use the
expressions given in Ref. \cite{noi3}. 
We notice that a host of configurations fall inside the region 
$R_m$; thus, the annual--modulation region is compatible with supersymmetric
configurations currently allowed by accelerator constraints.

\section{Interpretation of the annual--modulation data in terms of relic 
neutralinos}

Let us turn now to the implications of the DAMA/NaI experimental data, once 
these are interpreted in terms of relic neutralinos. 
The cosmological properties are examined first;
other prominent features of the relevant supersymmetric configurations 
will be examined afterwards.

\subsection{Neutralino cosmological properties}

For the derivation of the relic neutralino properties as regards its local 
density $\rho_{\chi}$ as well as its contribution to 
the average relic abundance $\Omega_{\chi} h^2$ compatible with region $R_m$, 
we adopt the 
straightforward procedure, outlined in Ref. \cite{noi2}, which does not require
any use of rescaling in the local neutralino density. The method is the
following: 

1) We evaluate $\sigma^{(\rm nucleon)}_{\rm scalar}$ and 
$\Omega_\chi h^2$ by varying the supersymmetric 
parameters over the grid and with the constraints defined in the previous 
section.

2) For any value of 
$[\rho_\chi \sigma^{(\rm nucleon)}_{\rm scalar}]_{\rm expt}$ 
compatible with the 
 region $R_m$ we calculate $\rho_\chi$ as given by 
$\rho_\chi = [\rho_\chi \sigma^{(\rm nucleon)}_{\rm scalar}]_{\rm expt} / 
\sigma^{(\rm nucleon)}_{\rm scalar}$
and restrict the values of $m_\chi$ to stay inside the region $R_m$ displayed
in Fig. 2. 

3) The results are then displayed in a scatter plot in the plane 
$\rho_\chi$ vs. $\Omega_\chi h^2$.

     Examples of our results are given in Figs. 4 a--c for a few 
experimentally allowed values of 
$\rho_{\chi} \sigma^{(\rm nucleon)}_{\rm scalar}$,
which bracket the range implied by the region $R_m$ of Fig. 2. 
Sects. a--c of Figs. 4 refer to the values 
$[\rho_{\chi}^{0.3} \sigma^{(\rm nucleon)}_{\rm scalar}]_{\rm expt} = 
4 \cdot 10^{-9}$ nbarn, 
$6 \cdot 10^{-9}$ nbarn and $8 \cdot 10^{-9}$ nbarn, respectively.

The two horizontal lines delimit the physical range  
0.1 GeV cm$^{-3} \leq \rho_l \leq $ 0.7 GeV cm$^{-3}$
for the total local density of non--baryonic dark matter. 
This (rather generous) range has been established by taking 
into account a possible 
flattening of the dark matter
halo  \cite{turner1,turner2}  and a possibly sizeable baryonic contribution to 
the galactic dark matter \cite{turner}.
The solid vertical lines  delimit the cosmologically 
interesting the range $0.01 \leq \Omega_{\chi} h^2 \leq 0.7$.  The two vertical
dashed lines delimit the range: 
$0.02 \leq \Omega_{\chi} h^2 \leq 0.2$, which represents the most appealing 
interval. 
Indeed, recent observations and
analyses \cite{omegamatter} favour 
$0.1 \lsim \Omega_{\rm matter} \lsim 0.4$. 
Combining this range with the one for $h$: 
$0.55 \lsim h \lsim 0.80$ \cite{hubble} and requiring that a 
cold dark matter candidate (such as the neutralino) supplies
$\sim (80$--$90)\%$ of $\Omega_{\rm matter}$, we actually obtain: 
$0.02 \lsim \Omega_{\rm CDM} h^2 \lsim 0.2$. 

The two slant 
dot--dashed lines delimit the band where linear rescaling procedure for the 
local density is usually applied. In Figs. 4 a--c
the upper dot--dashed line would refer to 
a rescaling with $(\Omega h^2)_{\rm min} = 0.01$, the lower one to 
the value $(\Omega h^2)_{\rm min} = 0.2$. However, notice that {\it in the
derivation of the scatter plot of Figs. 4 a--c, no use of rescaling 
for $\rho_{\chi}$ is made.}

     With the aid of this kind of plot we can classify the supersymmetric 
configurations belonging to region $R_m$ into various categories.
Configurations whose representative points fall above the maximum 
value $\rho_\chi = 0.7$ GeV cm$^{-3}$ 
have to be excluded (we remind that those providing 
an $\Omega_\chi h^2 > 0.7$ are already disregarded from the very beginning).  
Among the allowed configurations, those falling 
in the region inside 
both the  horizontal and solid vertical lines 
(called $A$ hereafter) are very 
appealing, since they would represent situations where the neutralino 
could have the role of a dominant cold dark matter component; even more so,
if the representative points fall 
in the subregion (called $B$ hereafter) inside the vertical 
band delimited by dashed lines. Configurations which fall inside 
the band delimited by the slant dot--dashed lines denote situations 
where the neutralino can only provide a fraction of the cold dark 
matter both at the level of local density and at the level of the 
average $\Omega$. Configurations above the upper dot--dashed line and below 
the upper horizontal solid line would imply a stronger 
clustering of neutralinos in our halo as compared to their average 
distribution in the Universe.

It is worth noticing a few important properties of 
the scatter plots shown in Figs. 4 a--c:

\begin{itemize}

\item [1)]
The scatter plots display a correlation between $\rho_\chi$ and 
$\Omega_\chi h^2$. This feature is expected on the basis of the 
following properties: i) $\Omega_\chi h^2$ is roughly inversely 
proportional to the neutralino pair annihilation cross section, 
ii) at fixed $[\rho_{\chi} \sigma^{(\rm nucleon)}_{\rm scalar}]_{\rm expt}$, 
$\rho_\chi$ is inversely proportional to 
$\sigma^{(\rm nucleon)}_{\rm scalar}$, 
iii) the annihilation cross section and 
$\sigma^{(\rm nucleon)}_{\rm scalar}$ are usually 
correlated functions (i.e., they are both increasing or 
decreasing functions of the supersymmetric parameters). 

\item [2)]  
The domains covered by the supersymmetric configurations in
regions $A$ and $B$ are slightly larger for smaller
values of 
$[\rho_{\chi} \sigma^{(\rm nucleon)}_{\rm scalar}]_{\rm expt}$.
This feature follows from the fact that 
$\sigma^{(\rm nucleon)}_{\rm scalar}$ is 
bounded from above by accelerator limits (mainly because of lower 
bounds on Higgs masses); this implies for $\rho_\chi$ 
a lower bound, which however is less stringent at lower values of 
$[\rho_{\chi} \sigma^{(\rm nucleon)}_{\rm scalar}]_{\rm expt}$.

\end{itemize}

We can conclude that the DAMA/NaI data are compatible with a relic 
neutralino as a major component of dark matter in the universe.

\subsection{Other properties of the supersymmetric configurations 
of set $S_m$}

Apart from the cosmological properties previously discussed, one of the most 
interesting questions is whether the supersymmetric configurations,
whose representative points fall inside the region $R_m$ of Fig. 2 
(hereafter denoted as configurations of set
$S_m$), are explorable at accelerators of the present 
generation. 
This point was investigated in Ref. \cite{noi3} in the case 
of the original DAMA/NaI annual--modulation region. Here we extend our 
considerations to the larger set $S_m$ derived from region $R_m$.

The supersymmetric configurations of set $S_m$ are constrained by 
the inequalities

\beq
\frac{[\rho_{\chi} \sigma_{\rm scalar}^{\rm (nucleon)}]_{\rm expt}}
{\rho_{\chi, {\rm max}}} \leq  \sigma_{\rm scalar}^{\rm (nucleon)} \leq 
\frac{[\rho_{\chi} \sigma_{\rm scalar}^{\rm (nucleon)}]_{\rm expt}}
{\rho_{\chi, {\rm min}}}
\label{eq:dis}
\eeq
where 
$\rho_{\chi,{\rm min}} = \xi \rho_{l,{\rm min}} = \xi \cdot$ 0.1 GeV cm$^{-3}$ 
and 
$\rho_{\chi,{\rm max}} = \xi \rho_{l,{\rm max}} = \xi \cdot$ 0.7 GeV cm$^{-3}$, 
and $[\rho_{\chi} \sigma_{\rm scalar}^{\rm (nucleon)}]_{\rm expt}$ is any value 
inside  region $R_m$ as a function of $m_{\chi}$. 

Eq.(\ref{eq:dis}) implies interesting correlations among the 
supersymmetric parameters, as displayed in Figs. 5--7. 
The first of these figures shows the scatter plot of $\tan \beta$ 
versus the mass $m_h$ of the lightest neutral CP-even Higgs boson 
(we recall that $m_h$ may be derived from the supersymmetric  parameters 
listed in Sect. III).  The correlation displayed in Fig. 5 between 
$\tan \beta$ and $m_h$  is implied by the fact that 
$\sigma_{\rm scalar}^{\rm (nucleon)}$ is usually dominated by 
Higgs-exchange amplitudes, and these are in turn large for large 
values of $\tan \beta$ and small values of $m_h$. 
As is apparent in Fig. 5 a number of supersymmetric 
configurations of set $S_m$ could still be explorable at LEP2, but the 
others require investigation at a high luminosity Fermilab Tevatron 
\cite{tev,baer} or at LHC. Fig. 6 shows 
a correlation between  $\tan \beta$ and the mass of the lightest top--squark 
$m_{\tilde t_1}$. The reasons for this feature are more involved, 
and are discussed in Ref. \cite{noi3}. 

A last plot, providing $\tan \beta$ versus $m_{\chi}$, is given in 
Fig. 7. Again, a correlation shows up here for not too large values of 
$\tan \beta$. LEP2 can only 
provide an investigation up to $m_{\chi} \simeq$ 50 GeV. The 
exploration up to $\sim$ 125 GeV can be performed with 
Tevatron upgrades (under favorable conditions), while for higher values of 
$m_{\chi}$ LHC is needed. 

\section{Conclusions}

We have investigated how the original DAMA/NaI annual--modulation region 
in the plane 
$\rho_{\chi}\sigma_{\rm scalar}^{\rm (nucleon)}$ -- $m_{\chi}$ 
is extended when the 
uncertainties in the galactic astrophysical velocities are taken into 
account.
One of the most noticeable consequences is that the range for the 
neutralino mass becomes 30 GeV $\lsim m_{\chi} \lsim$ 130 GeV at 1--$\sigma$ 
C.L., with a further
extension in the upper bound, when a possible bulk rotation of the dark matter 
halo is taken into account. However, the extension of the modulation region 
implies only slight modifications in the correlations among the supersymmetric
parameters in the MSSM scheme.

We have found that a number of supersymmetric configurations singled 
out  by the DAMA/NaI results have cosmological properties compatible with 
a relic neutralino as a dominant component of cold dark matter (on the 
average in our universe and in our galactic halo). 
It has also been discussed the discovery potential for the relevant 
supersymmetric configurations at accelerators of present generation.

\begin{center}
{\bf Acknowledgments}
\end{center}

A.B., F.D. and N.F. are grateful to Venya Berezinsky 
for interesting discussions about halo bulk rotations. 
This work was supported by DGICYT under grant number 
PB95--1077 and by the TMR network grant ERBFMRXCT960090 of 
the European Union.

\begin{center}
{\bf Note added}
\end{center}

During the completion of this work there appeared two preprints: 
hep-ph/9903467  (by L. Roskowsky) and 
hep-ph/9903468 (by M. Brhlik and L. Roszkowski), 
where some considerations on the effect of astrophysical
velocities uncertainty on WIMP direct searches are presented.

\newpage
\begin{table}[h]
\begin{center}
\caption{Results of the maximum likelihood method when varying the
velocity parameters ($v_0$ and $v_{\rm esc}$) 
of the WIMP velocity distribution. 
The values of $m_{\chi}$ and $\sigma_{\rm scalar}^{\rm (nucleon)}$ correspond to
the minima of the maximum likelihood method (with 1--$\sigma$ error bars).}
\begin{tabular}{|c|c|c|c|}
 $v_0$ & $v_{\rm esc}$ & $m_{\chi}$ & $\sigma_{\rm scalar}^{(\rm nucleon)}$  \\ 
 (km/s) & (km/s) & (GeV) & ($10^{-9}$ nbarn) \\ \hline
 220  & 450 & $59^{{+17}}_{-11}$  & $(7.3^{ +0.5}_{-1.2})$  \\ \hline
 220  & 550 & $59^{+16}_{-13}$  & $(7.0^{ +0.5}_{-1.2})$  \\ \hline
 220  & 650 & $59^{+17}_{-14}$  & $(7.0^{ +0.4}_{-1.2})$  \\ \hline
 170  & 550 & $95^{+29}_{-20}$  & $(8.3^{ +0.5}_{-1.2})$  \\ \hline
 270  & 550 & $41^{+14}_{-7} $  & $(6.8^{ +0.4}_{-1.3})$  
\end{tabular}
\end{center}
\end{table}

\newpage
\begin{figure}[t]
\hbox{
\psfig{figure=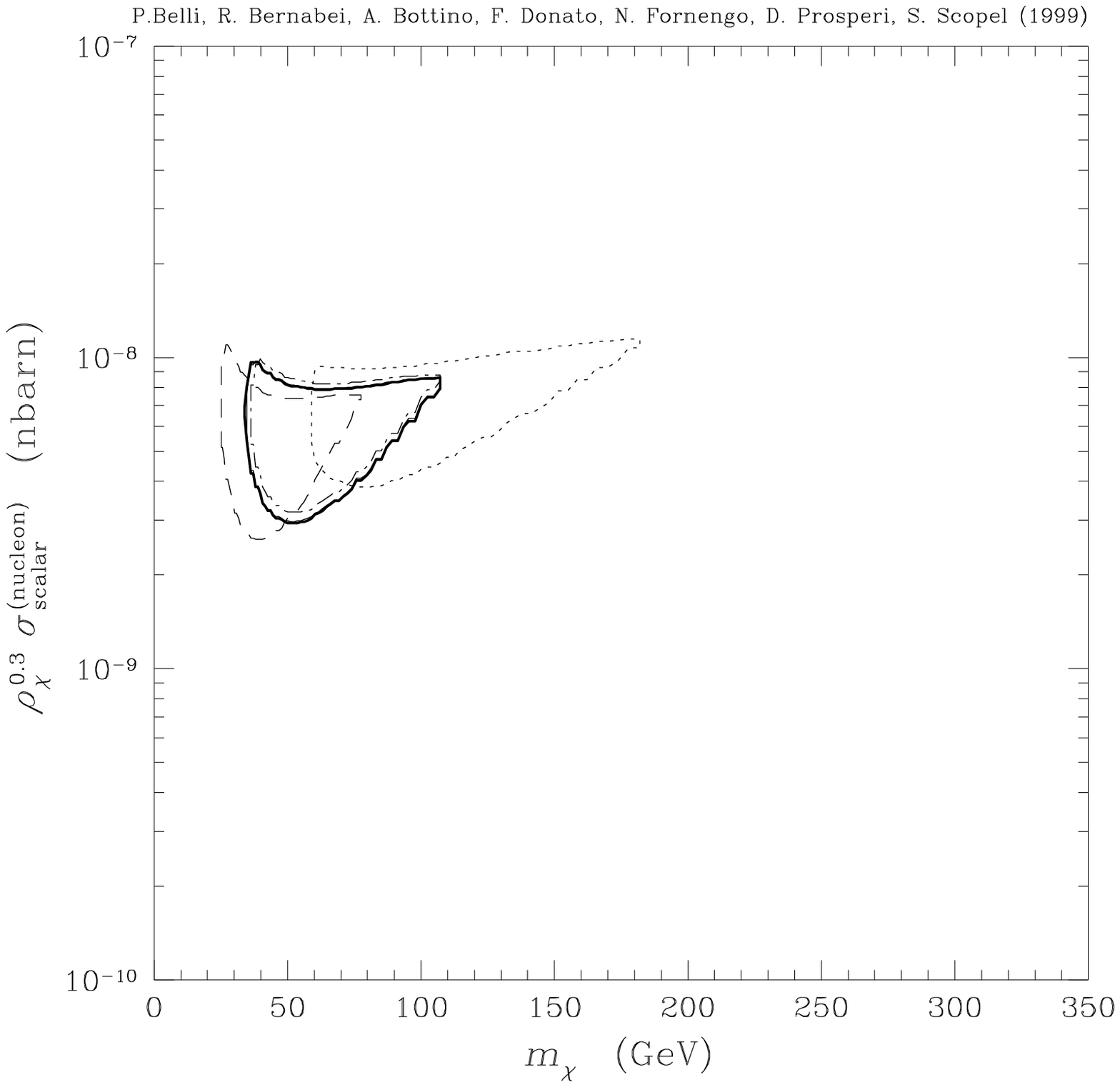,width=8.2in,bbllx=40bp,bblly=160bp,bburx=700bp,bbury=660bp,clip=}
}
{FIG. 1. Annual--modulation regions singled out by the DAMA/NaI 
experiment [3] in the plane 
$\rho_{\chi}^{0.3} \sigma_{\rm scalar}^{\rm (nucleon)} - m_{\chi}$ for 
various values of velocities $v_0$ and $v_{esc}$:
$v_0 = 220$ km s$^{-1}$, $v_{esc} = 650$ km s$^{-1}$ (solid line); 
$v_0 = 220$ km s$^{-1}$, $v_{esc} = 450$ km s$^{-1}$ (dot--dashed line);
$v_0 = 170$ km s$^{-1}$, $v_{esc} = 550$ km s$^{-1}$ (dotted line); 
$v_0 = 270$ km s$^{-1}$, $v_{esc} = 550$ km s$^{-1}$ (dashed line).
The countour line for
$v_0 = 220$ km s$^{-1}$, $v_{esc} = 550$ km s$^{-1}$ is undistinguishable
on the plot from the solid line.
The quantity $\rho_{\chi}^{0.3}$ is the local neutralino matter
density in units of 0.3 GeV cm$^{-3}$, i.e.
$\rho_{\chi}^{0.3} \def \rho_{\chi}$/(0.3 GeV cm$^{-3}$).
}
\end{figure}

\newpage
\begin{figure}[t]
\hbox{
\psfig{figure=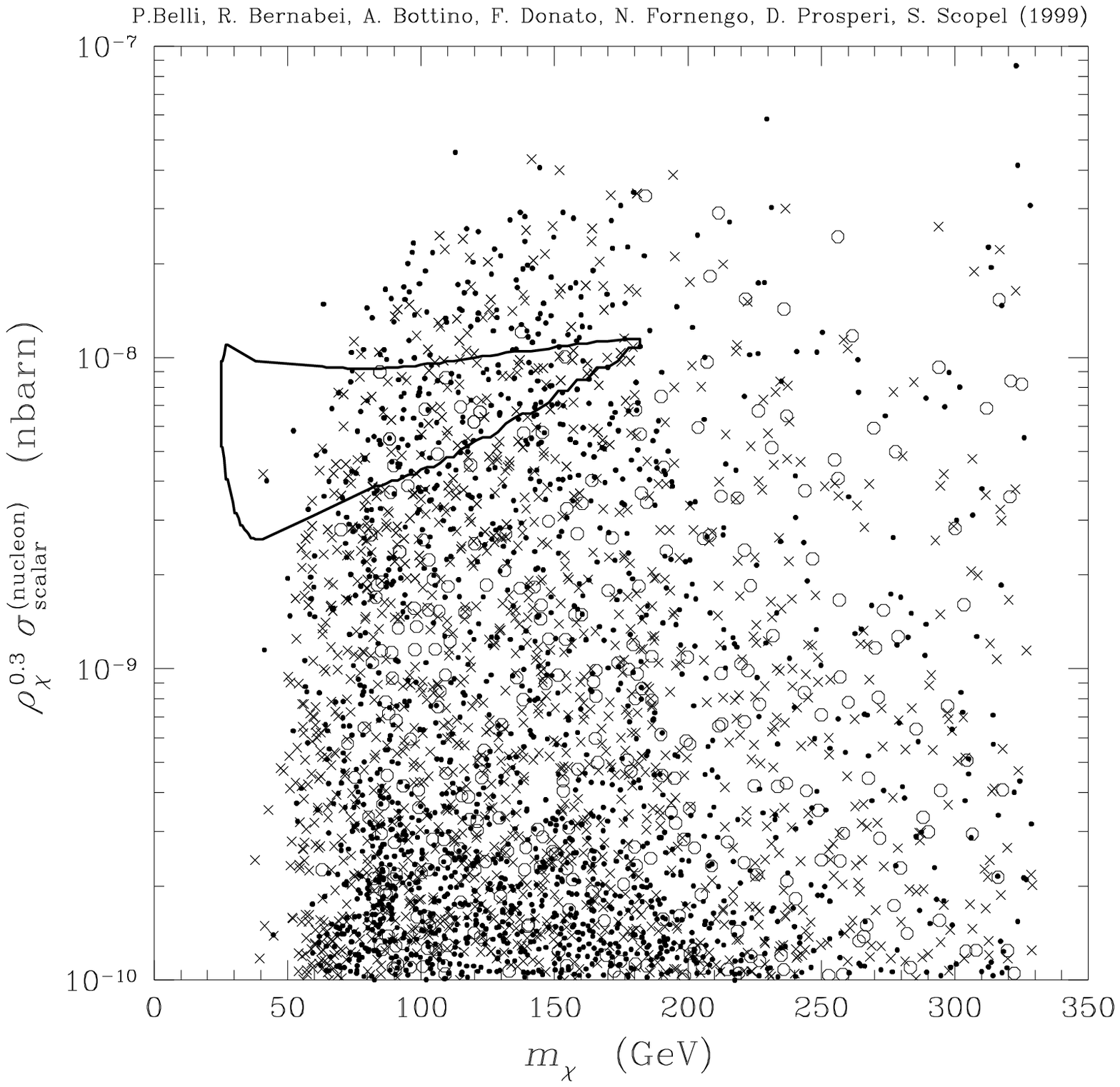,width=8.2in,bbllx=40bp,bblly=160bp,bburx=700bp,bbury=660bp,clip=}
}
{FIG. 2. The contour line delimits the 
annual--modulation region obtained by varying the velocities $v_0$ and 
$v_{esc}$ over the ranges given in Eqs. (\ref{eq:not} -- \ref{eq:esc}). The 
scatter 
plot represents the theoretical predictions of a generic MSSM, as  
described in Sect. III. Different symbols identify different neutralino
compositions: circles stand for a higgsino, crosses for a gaugino
and dots for a mixed neutralino. 
}
\end{figure}

\newpage
\begin{figure}[t]
\hbox{
\psfig{figure=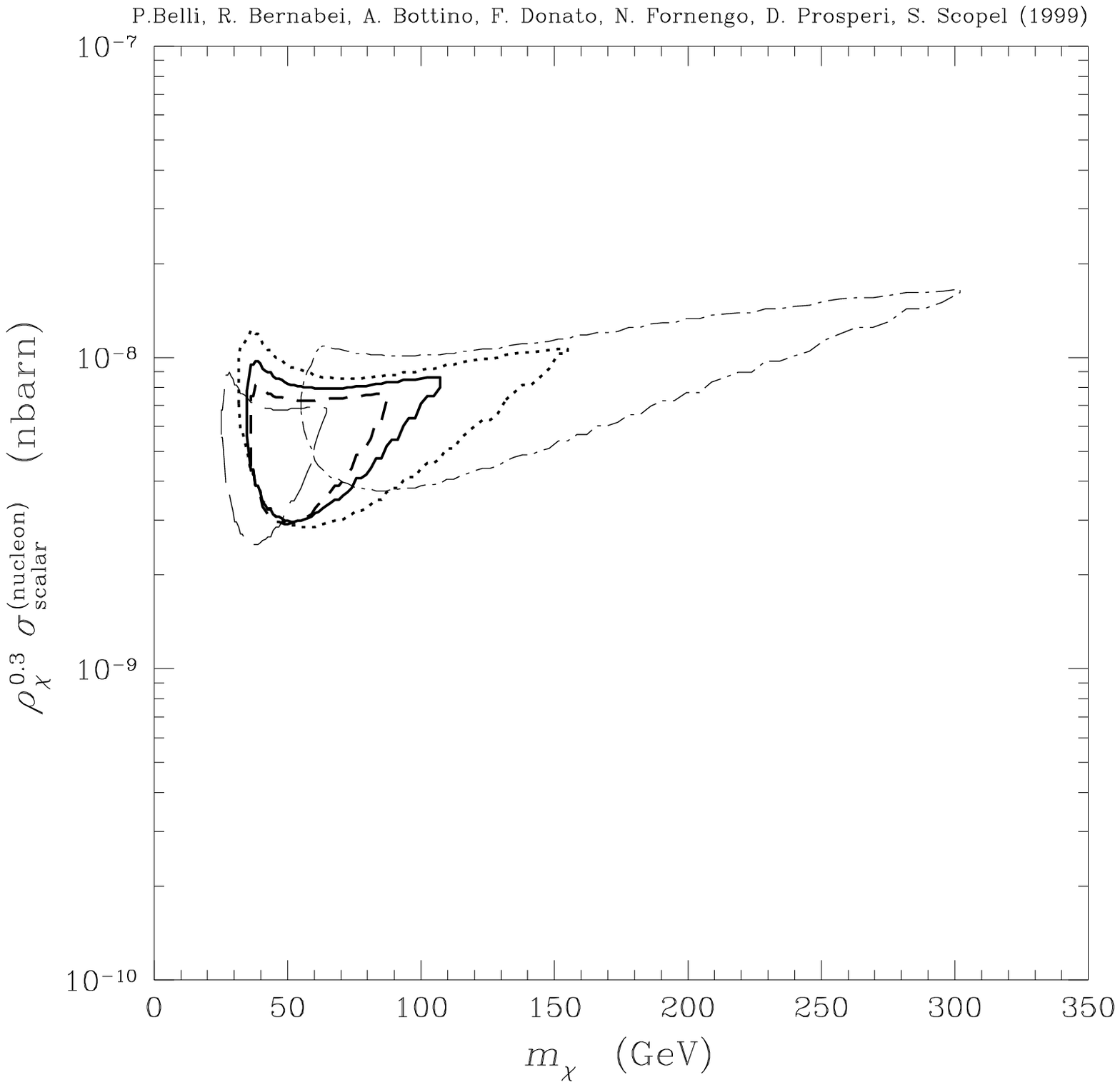,width=8.2in,bbllx=40bp,bblly=160bp,bburx=700bp,bbury=660bp,clip=}
}
{FIG. 3. Annual--modulation regions when a possible bulk rotation of the
dark matter halo is taken into account. The contour lines
refer to the following halo parameters:
$v_0 = 220$ km s$^{-1}$, non rotating halo (solid line); 
$v_0 = 220$ km s$^{-1}$, co--rotating halo with spin parameter
	$\lambda = 0.05$ (dotted line);
$v_0 = 220$ km s$^{-1}$, counter--rotating halo with
	$\lambda = 0.05$ (dashed line);
$v_0 = 170$ km s$^{-1}$, co--rotating halo with
	$\lambda = 0.05$ (dot--dashed line);
$v_0 = 270$ km s$^{-1}$, counter--rotating halo with
	$\lambda = 0.05$ (long--dashed line);
The escape velocity is fixed at the value 
$v_{esc} = 550$ km s$^{-1}$ for all
the contours. 
}
\end{figure}

\newpage
\begin{figure}[t]
\hbox{
\psfig{figure=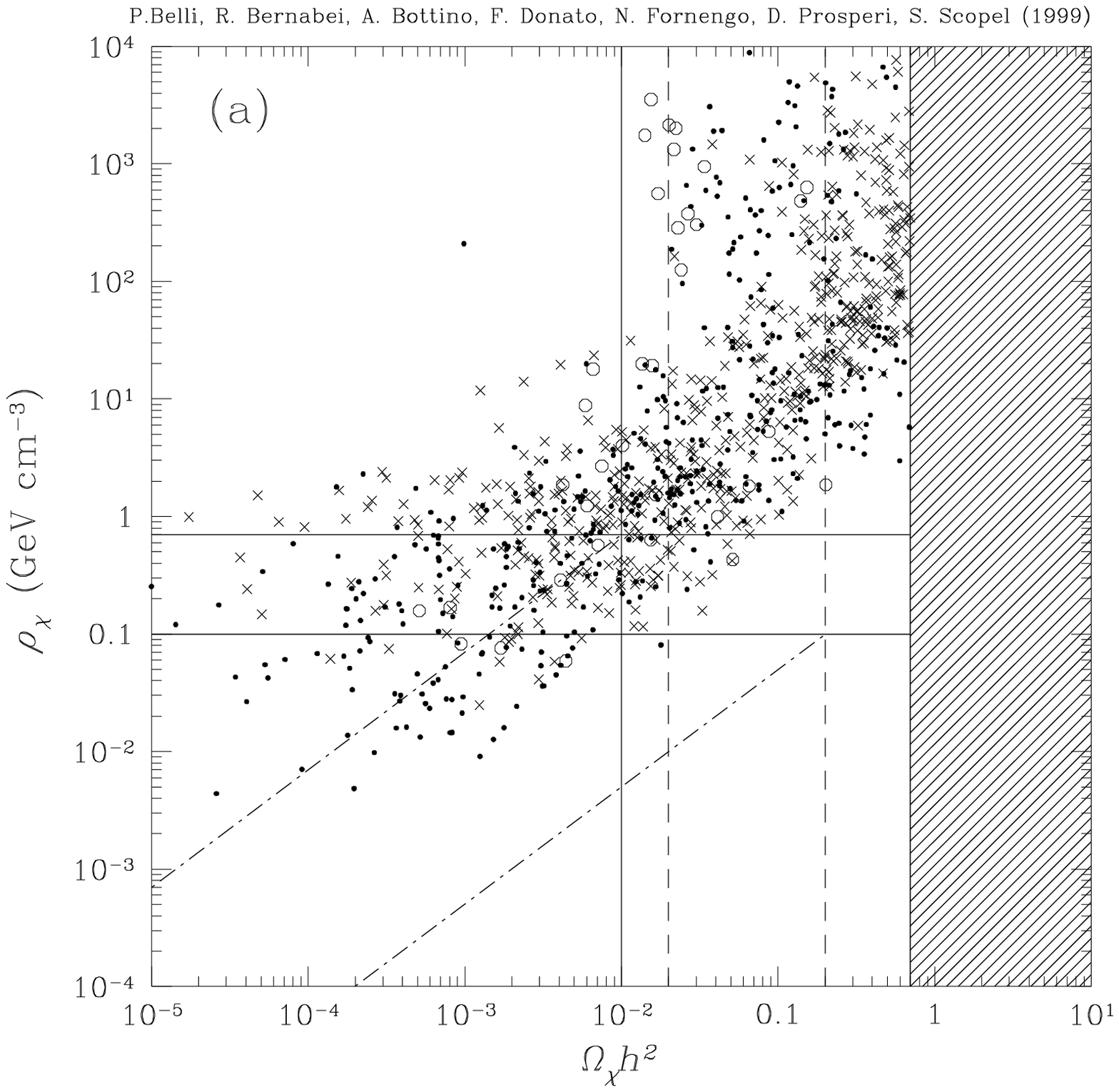,width=8.2in,bbllx=40bp,bblly=160bp,bburx=700bp,bbury=660bp,clip=}
}
{FIG. 4a. The neutralino local density $\rho_{\chi}$, calculated for  fixed 
values of 
$[\rho_{\chi}^{0.3} \sigma_{\rm scalar}^{\rm nucleon}]_{expt}$, is plotted 
versus the neutralino relic abundance 
$\Omega_{\chi} h^2$. For any value of 
$[\rho_{\chi}^{0.3} \sigma_{\rm scalar}^{\rm nucleon}]_{expt}$
the neutralino mass is restricted to the  range dictated by the 
contour line of Fig. 2. The different panels of the figure 
correspond to the following choices: 
(a) $[\rho_{\chi}^{0.3} \sigma_{\rm scalar}^{\rm nucleon}]_{expt}$=
    $ 4\cdot 10^{-9}$ nbarn and 28 GeV $\leq m_\chi \leq$ 88 GeV;
(b) $[\rho_{\chi}^{0.3} \sigma_{\rm scalar}^{\rm nucleon}]_{expt}$=
    $ 6\cdot 10^{-9}$ nbarn and 25 GeV $\leq m_\chi \leq$ 131 GeV;
(c) $[\rho_{\chi}^{0.3} \sigma_{\rm scalar}^{\rm nucleon}]_{expt}$=
    $ 8\cdot 10^{-9}$ nbarn and 25 GeV $\leq m_\chi \leq$ 156 GeV.
The two horizontal lines delimit the physical range for the local 
density for non-baryonic dark matter. The two solid vertical lines delimit 
the interval of $\Omega_{\chi} h^2$ of cosmological interest. The 
two vertical dashed lines delimit the  preferred band for cold 
dark matter. The two slant  dot--dashed lines delimit the 
band where linear rescling procedure is ususally applied. 
}
\end{figure}

\newpage
\begin{figure}[t]
\hbox{
\psfig{figure=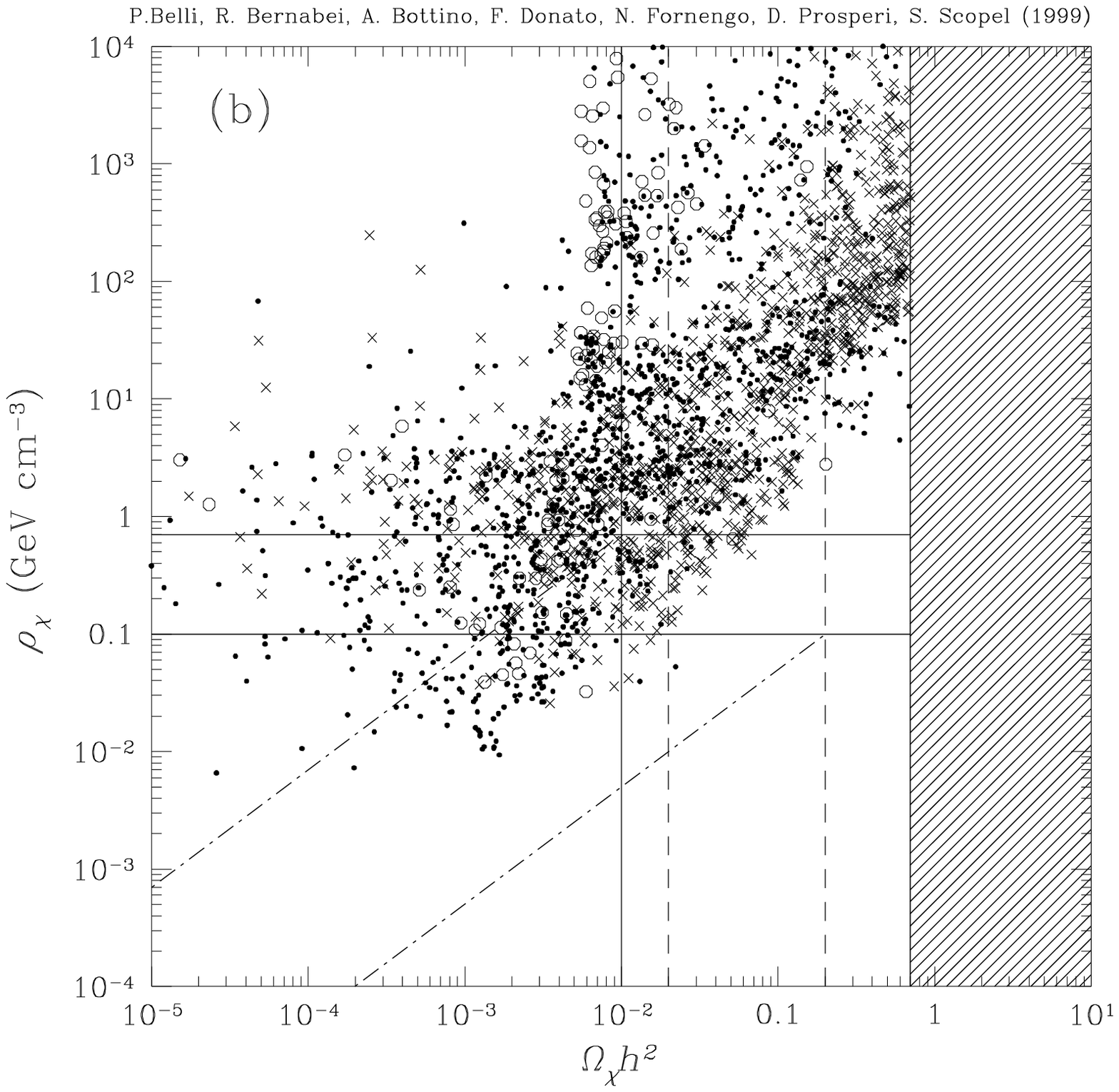,width=8.2in,bbllx=40bp,bblly=160bp,bburx=700bp,bbury=660bp,clip=}
}
{FIG. 4b. The neutralino local density $\rho_{\chi}$, calculated for  fixed 
values of 
$[\rho_{\chi}^{0.3} \sigma_{\rm scalar}^{\rm nucleon}]_{expt}$, is plotted 
versus the neutralino relic abundance 
$\Omega_{\chi} h^2$. For any value of 
$[\rho_{\chi}^{0.3} \sigma_{\rm scalar}^{\rm nucleon}]_{expt}$
the neutralino mass is restricted to the  range dictated by the 
contour line of Fig. 2. The different panels of the figure 
correspond to the following choices: 
(a) $[\rho_{\chi}^{0.3} \sigma_{\rm scalar}^{\rm nucleon}]_{expt}$=
    $ 4\cdot 10^{-9}$ nbarn and 28 GeV $\leq m_\chi \leq$ 88 GeV;
(b) $[\rho_{\chi}^{0.3} \sigma_{\rm scalar}^{\rm nucleon}]_{expt}$=
    $ 6\cdot 10^{-9}$ nbarn and 25 GeV $\leq m_\chi \leq$ 131 GeV;
(c) $[\rho_{\chi}^{0.3} \sigma_{\rm scalar}^{\rm nucleon}]_{expt}$=
    $ 8\cdot 10^{-9}$ nbarn and 25 GeV $\leq m_\chi \leq$ 156 GeV.
The two horizontal lines delimit the physical range for the local 
density for non-baryonic dark matter. The two solid vertical lines delimit 
the interval of $\Omega_{\chi} h^2$ of cosmological interest. The 
two vertical dashed lines delimit the  preferred band for cold 
dark matter. The two slant  dot--dashed lines delimit the 
band where linear rescling procedure is ususally applied. 
}
\end{figure}

\newpage
\begin{figure}[t]
\hbox{
\psfig{figure=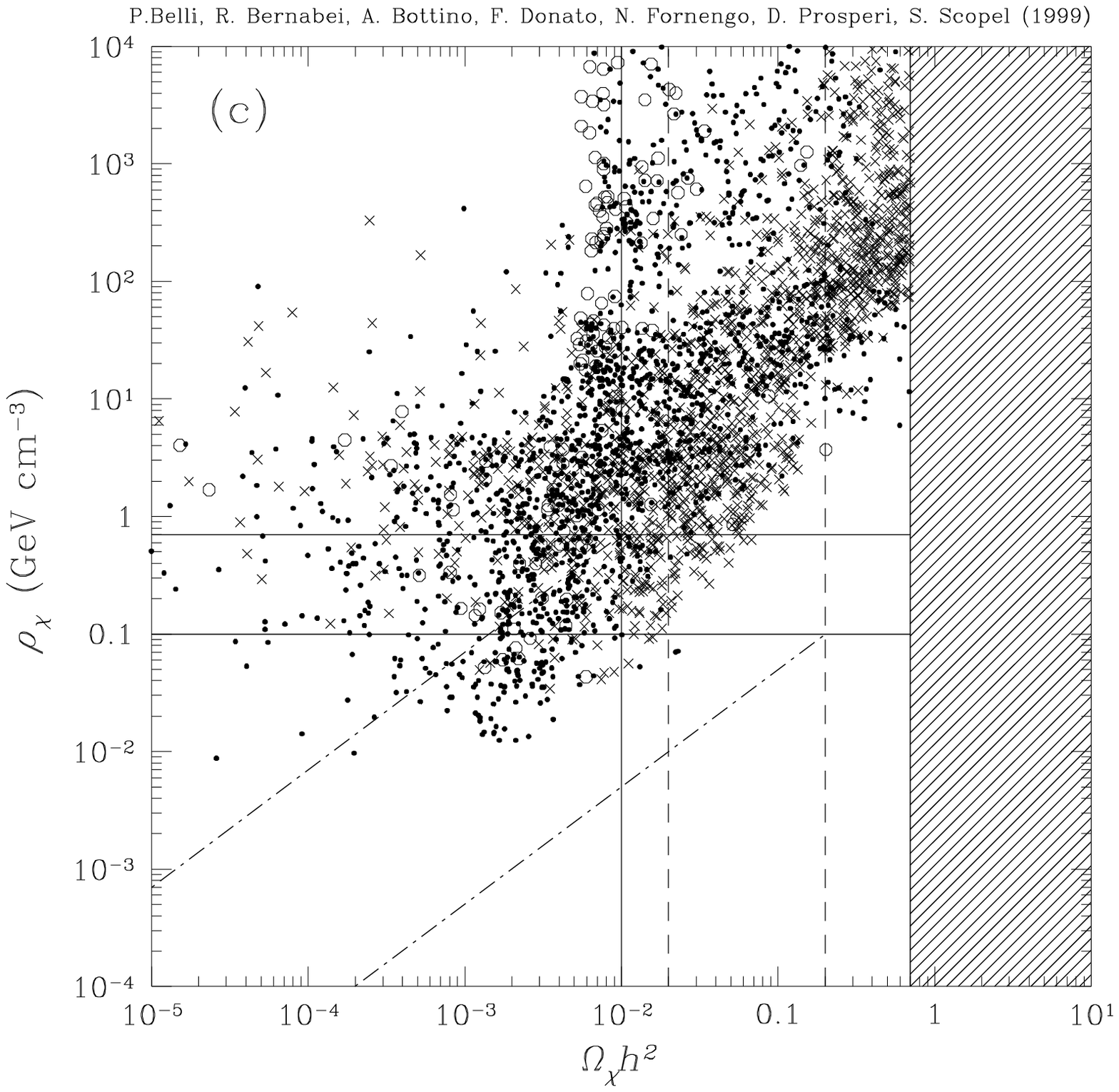,width=8.2in,bbllx=40bp,bblly=160bp,bburx=700bp,bbury=660bp,clip=}
}
{FIG. 4c. The neutralino local density $\rho_{\chi}$, calculated for  fixed 
values of 
$[\rho_{\chi}^{0.3} \sigma_{\rm scalar}^{\rm nucleon}]_{expt}$, is plotted 
versus the neutralino relic abundance 
$\Omega_{\chi} h^2$. For any value of 
$[\rho_{\chi}^{0.3} \sigma_{\rm scalar}^{\rm nucleon}]_{expt}$
the neutralino mass is restricted to the  range dictated by the 
contour line of Fig. 2. The different panels of the figure 
correspond to the following choices: 
(a) $[\rho_{\chi}^{0.3} \sigma_{\rm scalar}^{\rm nucleon}]_{expt}$=
    $ 4\cdot 10^{-9}$ nbarn and 28 GeV $\leq m_\chi \leq$ 88 GeV;
(b) $[\rho_{\chi}^{0.3} \sigma_{\rm scalar}^{\rm nucleon}]_{expt}$=
    $ 6\cdot 10^{-9}$ nbarn and 25 GeV $\leq m_\chi \leq$ 131 GeV;
(c) $[\rho_{\chi}^{0.3} \sigma_{\rm scalar}^{\rm nucleon}]_{expt}$=
    $ 8\cdot 10^{-9}$ nbarn and 25 GeV $\leq m_\chi \leq$ 156 GeV.
The two horizontal lines delimit the physical range for the local 
density for non-baryonic dark matter. The two solid vertical lines delimit 
the interval of $\Omega_{\chi} h^2$ of cosmological interest. The 
two vertical dashed lines delimit the  preferred band for cold 
dark matter. The two slant  dot--dashed lines delimit the 
band where linear rescling procedure is ususally applied. 
}
\end{figure}

\newpage
\begin{figure}[t]
\hbox{
\psfig{figure=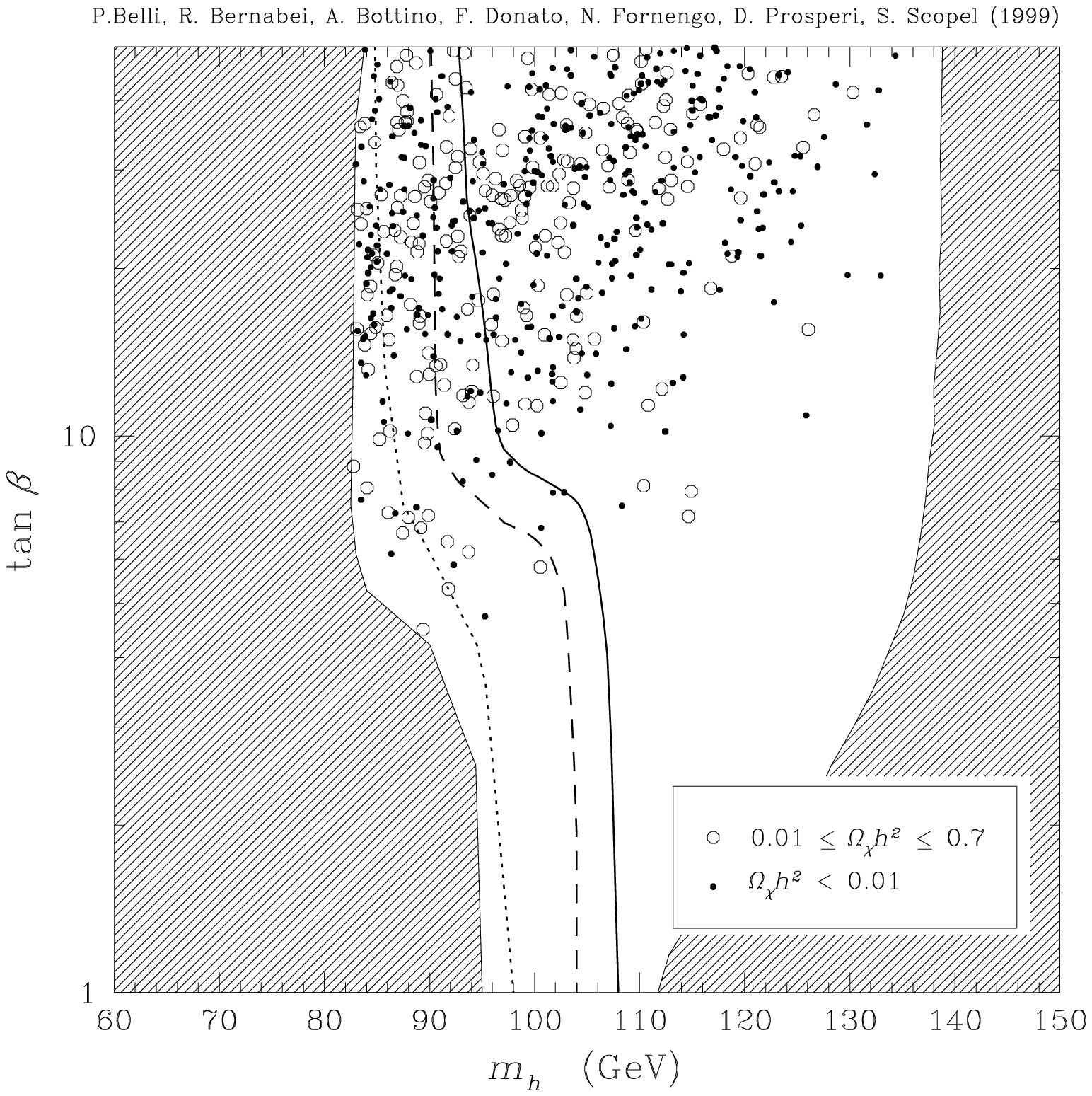,width=8.2in,bbllx=40bp,bblly=160bp,bburx=700bp,bbury=660bp,clip=}
}
{FIG. 5. Scatter plot for set $S$ in the plane $m_h$ -- $\tan \beta$. 
The hatched region on the right is excluded by theory. 
The hatched region on the left is 
excluded by present LEP data at $\sqrt s$ = 189 GeV. The dotted and the dashed 
curves denote the reach of LEP2 at energies $\sqrt s$ = 192 GeV and 
$\sqrt s$ = 200 GeV, respectively. The solid line represents the 
95\% C.L. bound reachable at LEP2, in case of non discovery of a neutral 
Higgs boson.
}
\end{figure}

\newpage
\begin{figure}[t]
\hbox{
\psfig{figure=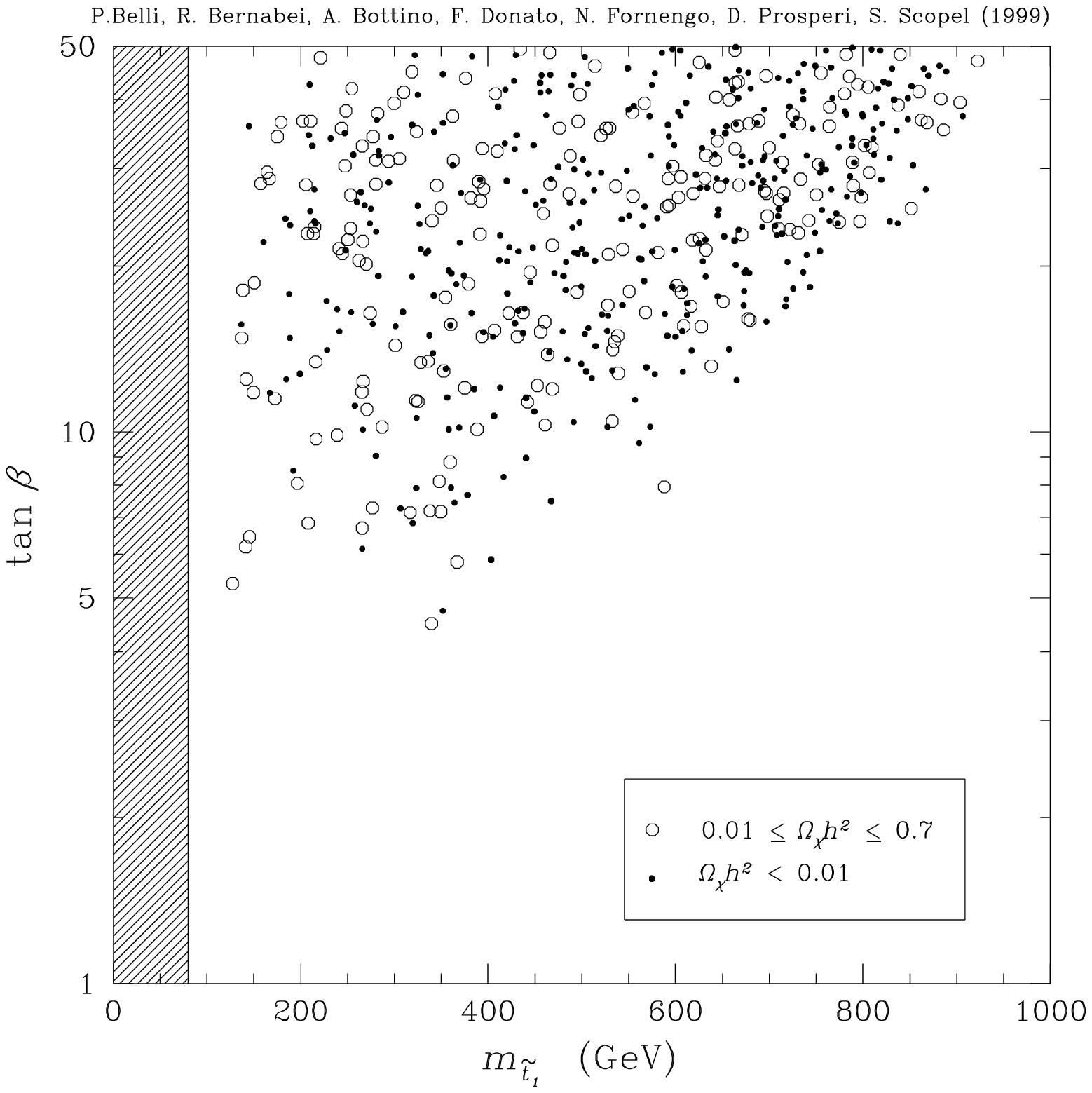,width=8.2in,bbllx=40bp,bblly=160bp,bburx=700bp,bbury=660bp,clip=}
}
{FIG. 6. Scatter plot for set $S$ in the plane 
$m_{\tilde t_1}$ -- $\tan \beta$.  
The hatched region is excluded by LEP data.  
}
\end{figure}

\newpage
\begin{figure}[t]
\hbox{
\psfig{figure=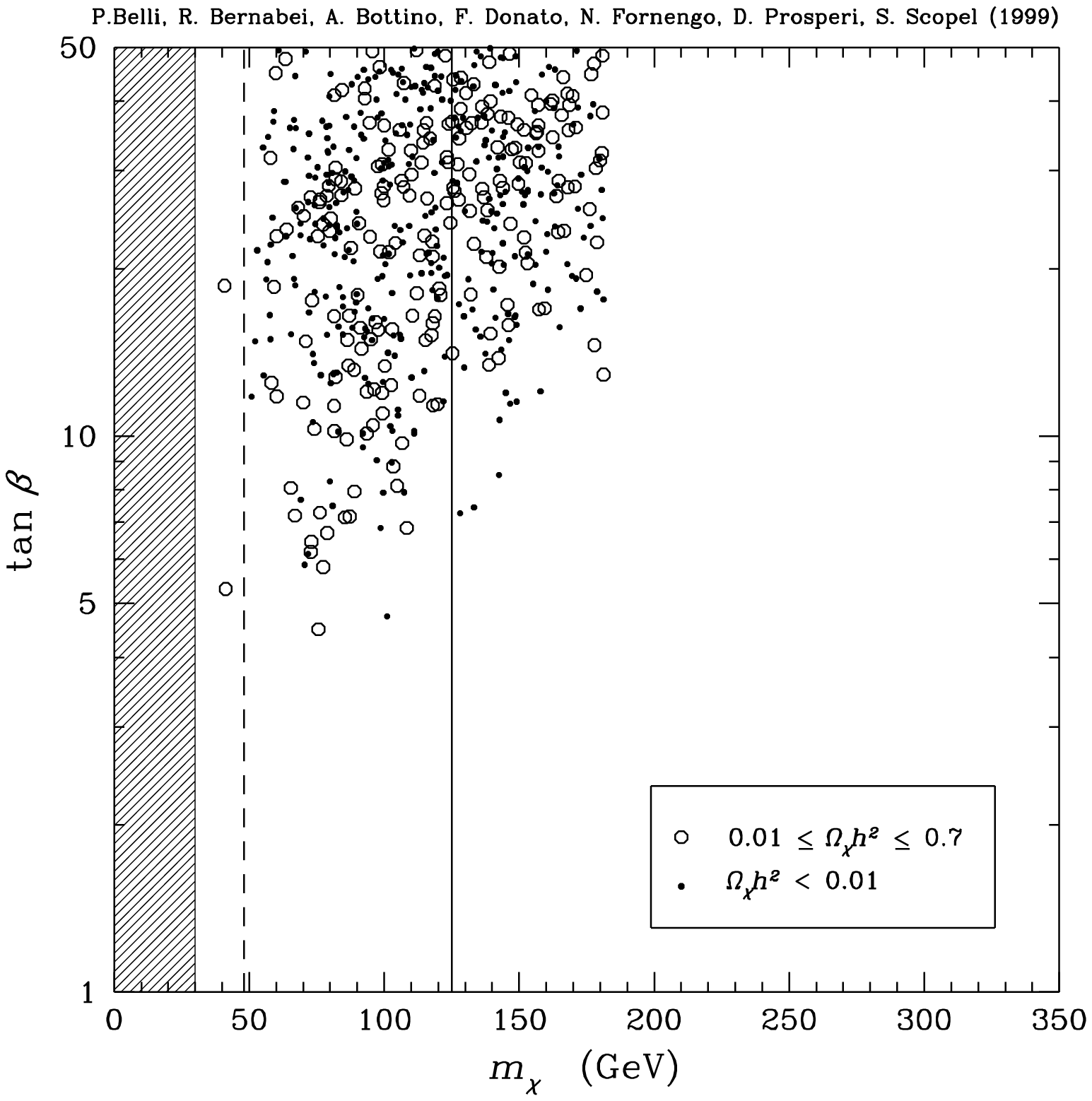,width=8.2in,bbllx=40bp,bblly=160bp,bburx=700bp,bbury=660bp,clip=}
}
{FIG. 7.  Scatter plot for set $S$ in the plane 
$m_{\chi}$ -- $\tan \beta$. 
The hatched region on the left is 
excluded by present LEP data. The dashed and the 
solid vertical lines denote the reach of 
LEP2 and TeV33, respectively. 
}
\end{figure}

\end{document}